\documentclass[twocolumn]{aastex63}

\usepackage{multirow}

\newcommand{\xmm} {{\it XMM-Newton}}
\newcommand{\chandra} {{\it Chandra}}
\newcommand{\nustar} {{\it NuSTAR}}
\newcommand{\swift} {{\it Swift}}

\newcommand{\swiftxrt} {{\it Swift}/XRT}

\newcommand{\gaia} {{\it Gaia}}

\newcommand{\cmsq} {cm$^{-2}$}
\newcommand{\nh} {$N_{\rm{H}}$}
\newcommand{\lx} {$L_{\rm{X}}$}
\newcommand{\fx} {$F_{\rm{X}}$}

\newcommand{\fexxv}{{\rm{Fe\,\sc{xxv}}}}
\newcommand{\fexxvi}{{\rm{Fe\,\sc{xxvi}}}}

\newcommand{\oviii}{{\rm{O\,\sc{viii}}}}
\newcommand{\nex}{{\rm{Ne\,\sc{x}}}}
\newcommand{\sixiv}{{\rm{Si\,\sc{xiv}}}}
\newcommand{\sxvi}{{\rm{S\,\sc{xvi}}}}

\newcommand{\degree}{{$^\circ$}}
\newcommand{\ergs}{\mbox{\thinspace erg\thinspace s$^{-1}$}}
\newcommand{\ergcms}{\mbox{\thinspace erg\thinspace cm$^{-2}$\thinspace s$^{-1}$}}

\newcommand{\cntrt}{counts\,s$^{-1}$}
\newcommand{\kms}{km\,s$^{-1}$}

\newcommand{\msol} {$M_{\odot}$}

\shorttitle{A UFO or CRSF in NGC 4045 ULX?}
\shortauthors{Brightman et al.}

\begin{document}

\title{An 8.56 keV absorption line in the hyperluminous X-ray source in NGC 4045: ultra-fast outflow or cyclotron line?}

\author{Murray Brightman}
\affiliation{Cahill Center for Astrophysics, California Institute of Technology, 1216 East California Boulevard, Pasadena, CA 91125, USA}

\author{Peter Kosec}
\affiliation{MIT Kavli Institute for Astrophysics and Space Research, Cambridge, MA 02139, USA}

\author{Felix F\"{u}rst}
\affiliation{Quasar Science Resources SL for ESA, European Space Astronomy Centre (ESAC), Science Operations Departement, 28692 Villanueva de la Cañada, Madrid, Spain}

\author{Hannah Earnshaw}
\affiliation{Cahill Center for Astrophysics, California Institute of Technology, 1216 East California Boulevard, Pasadena, CA 91125, USA}

\author{Marianne Heida}
\affiliation{European Southern Observatory, Garching, Germany}

\author{Matthew J Middleton}
\affiliation{Department of Physics \& Astronomy, University of Southampton, Southampton, SO17 1BJ, UK}

\author{Daniel Stern}
\affiliation{Jet Propulsion Laboratory, California Institute of Technology, Pasadena, CA 91109, USA}

\author{Dominic J Walton}
\affiliation{Centre for Astrophysics Research, University of Hertfordshire, College Lane, Hatfield AL10 9AB, UK}

\email{murray@srl.caltech.edu}

\begin{abstract}

We report on the discovery of an absorption line at $E=8.56^{+0.05}_{-0.11}$ keV detected with a significance of $>3.3\sigma$ in the \nustar\ and \xmm\ spectra of a newly discovered hyperluminous X-ray source (HLX, \lx$>10^{41}$ \ergs) in the galaxy NGC 4045 at a distance of 32 Mpc. The source was first discovered serendipitously in a \swiftxrt\ observation of the galaxy, and \swift\ monitoring reveals a highly variable source changing by over an order of magnitude from maximum to minimum. The origin of the absorption line appears likely to be by highly ionized iron with a blue shift of 0.19$c$, indicating an ultrafast outflow (UFO). However, the large equivalent width of the line (EW$=-0.22^{+0.08}_{-0.09}$ keV) paired with the lack of other absorption lines detected are difficult to reconcile with models. An alternative explanation is that the line is due to a cyclotron resonance scattering feature (CRSF), produced by the interaction of X-ray photons with the powerful magnetic field of a neutron star.

\end{abstract}


\section{Introduction}

Ultraluminous X-ray sources (ULXs) are X-ray sources located outside the nucleus of galaxies with observed fluxes that imply isotropic luminosities greater than 10$^{39}$ \ergs. This luminosity is equivalent to the Eddington luminosity of a 10 \msol\ black hole, the typical mass of known stellar-remnant black holes in our Galaxy. Hence ULXs are either shining at super-Eddington rates, or the mass of the compact object is greater than 10 \msol. They are therefore of interest for studies of extreme accretion and black hole demographics.

Hyperluminous X-ray sources (HLXs) constitute the extreme end of the ULX luminosity function, with luminosities greater than $10^{41}$ \ergs, and are rare, with only 71 out of 1843 (4\%) ULX candidates listed in the latest ULX catalog of \cite{walton21} reaching this luminosity. Their luminosities would seem to imply a black hole mass of $>1000$ \msol, an {\it intermediate mass black hole} (IMBH); or a 10 \msol\ black hole shining at 100 times Eddington. ESO 243-49 HLX-1 \citep{farrell09} is a well known HLX thought to be powered by an IMBH, however, another HLX, NGC 5907 ULX1 is known to be powered by a neutron star due to the detection of X-ray pulsations \citep{israel17a}. With a mass of only 1--2 \msol, the implied luminosity is $\sim500$ times its Eddington luminosity. Several other ULX pulsars are also known, which include M82 X-2 \citep{bachetti14}, NGC~5907~ULX \citep{israel17a}, NGC~7793~P13 \citep{israel17,fuerst17}, NGC~300~ULX \citep{carpano18}, NGC~1313~X-2 \citep{sathyaprakash19}, and M51 ULX7 \citep{rodriguez20}. 

At such high apparent super-Eddington luminosities, powerful radiation driven winds are expected from ULXs \citep{shakura73,poutanen07}. While no signatures for these outflows have been detected in any HLX source to date, likely due to their small numbers, such outflows have been detected in a number of ULXs, such as NGC 1313 X-1, NGC 5408 X-1, and NGC 6946 X-1 \citep{pinto16,pinto20}. These have mostly been detected with the reflection grating spectrometer (RGS) instrument on \xmm, e.g. NGC 55 ULX \citep{pinto17}, NGC 5204 X-1 \citep{kosec18a} and NGC 300 ULX1 \citep{kosec18}. Evidence for these was first seen in \xmm/pn data as soft X-ray residuals \citep{middleton14}. \cite{walton16} also found evidence for the outflow from NGC 1313 X-1 in \xmm/pn and \nustar\ data at 8.77 keV and \cite{kosec18} found evidence for the outflow from NGC 300 ULX1 in \xmm/pn data.

In addition to these atomic absorption lines in ULX spectra, \cite{brightman18} reported the detection of a strong absorption line at 4.5 keV in the \chandra/ACIS spectrum of ULX8 in M51. Since the energy of this line was not consistent with atomic absorption, the authors concluded that it was due to a cyclotron resonance scattering feature (CRSF), produced by the interaction of X-ray photons with a powerful magnetic field \citep{gnedin74,truemper78}. This naturally identified the accretor as a neutron star, since black holes cannot produce such strong magnetic fields. \cite{walton18b} also identified a potential CRSF at 13 keV in the pulsed \nustar\ spectrum of NGC 300 ULX1. Therefore, detecting and studying absorption features in the X-ray spectra of ULXs can reveal important information about the compact object powering the source, and the extreme accretion onto it.

Here we report the discovery of a new HLX candidate in NGC 4045 with \swiftxrt, and subsequent observations with \chandra, \nustar\ and \xmm. NGC 4045 is a spiral galaxy at a distance of 32.1 Mpc as determined from the Tully-Fisher relation \citep{tully16} with a redshift of $z=0.00659$, and hosts an optically identified AGN \citep{gavazzi11}. The galaxy also hosted the type II supernova SN 1985B \citep{kosai85}. Uncertainties are given at the 90\% confidence level unless otherwise stated.

\section{X-ray Data Analysis}

\subsection{Swift}

\begin{figure}[h]
\begin{center}
\includegraphics[trim=10 20 20 0, width=80mm]{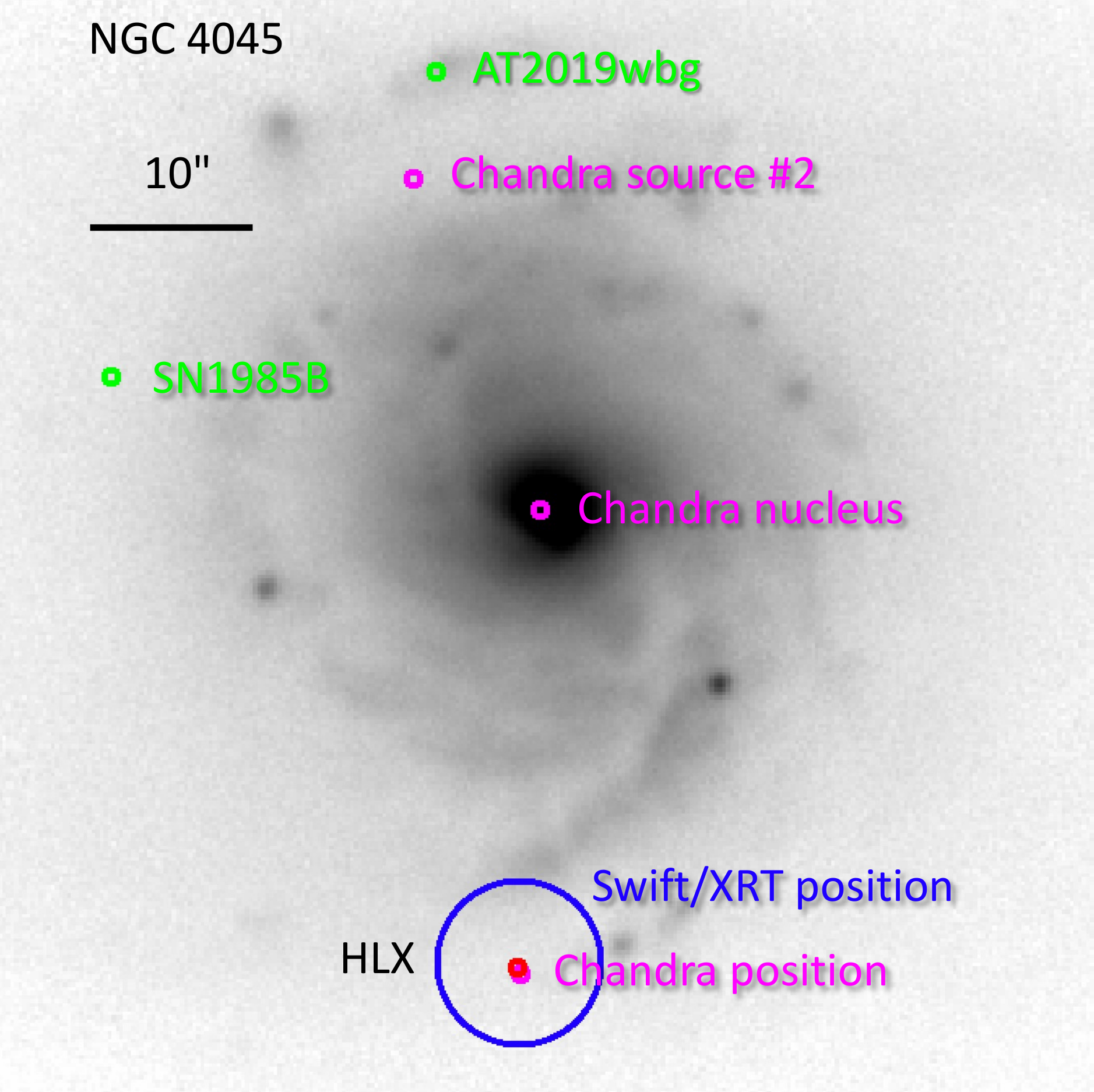}
\end{center}
\caption{PanSTARRS $r$-band image of NGC 4045. The position of the X-ray source detected by \swiftxrt\ is shown with a blue circle where the radius represents the positional uncertainty. The more accurate position provided by \chandra\ is shown with a magenta circle. This appears to place the source in one of the spiral arms of the galaxy. The positions of SN1985B and AT2019wbg are also shown, along with the position of the nucleus and another X-ray source detected in the galaxy by \chandra. }
\label{fig_ngc4045_img}
\end{figure}

We have been searching for new X-ray sources in observations made by NASA's {\it Neil Gehrels Swift Observatory} \citep{gehrels04}, specifically using its X-ray Telescope \citep[XRT,][]{burrows05}. This search has already uncovered an X-ray luminous tidal disruption event \citep{brightman21}.
On 2019 December 4, we detected a source in an observation of AT2019wbg (obsID 00012842001), a candidate supernova hosted by NGC 4045. This was done with the {\tt detect} function of the {\sc heasoft} tool {\sc ximage} using a signal to noise threshold of 3. The X-ray source was 55\arcsec\ from AT2019wbg and therefore not related. We used the online tool provided by the University of Leicester\footnote{https://www.swift.ac.uk/user\_objects/} \citep{evans07,evans09} to obtain the best position of the source, which gave R.A. = 12h 02m 42.360s (180.6765\degree), Decl.=+1\degree\ 58\arcmin\ 08.54\arcsec\ (1.9690389\degree, J2000), with an uncertainty of 5.1\arcsec\ (90\% confidence). This placed the source outside of the nucleus of NGC 4045, and in one of its spiral arms (Figure \ref{fig_ngc4045_img}). No X-ray source had been reported at this position previously. This included two \swift\ observations taken only 1 month prior to its initial detection as part of the \swift\ Gravitational Wave Galaxy Survey \citep[SGWGS,][]{klingler19}. The position of the source had not been previously observed with \chandra, \xmm, or \nustar.

\begin{figure}[h]
\begin{center}
\includegraphics[trim=10 20 20 0, width=95mm]{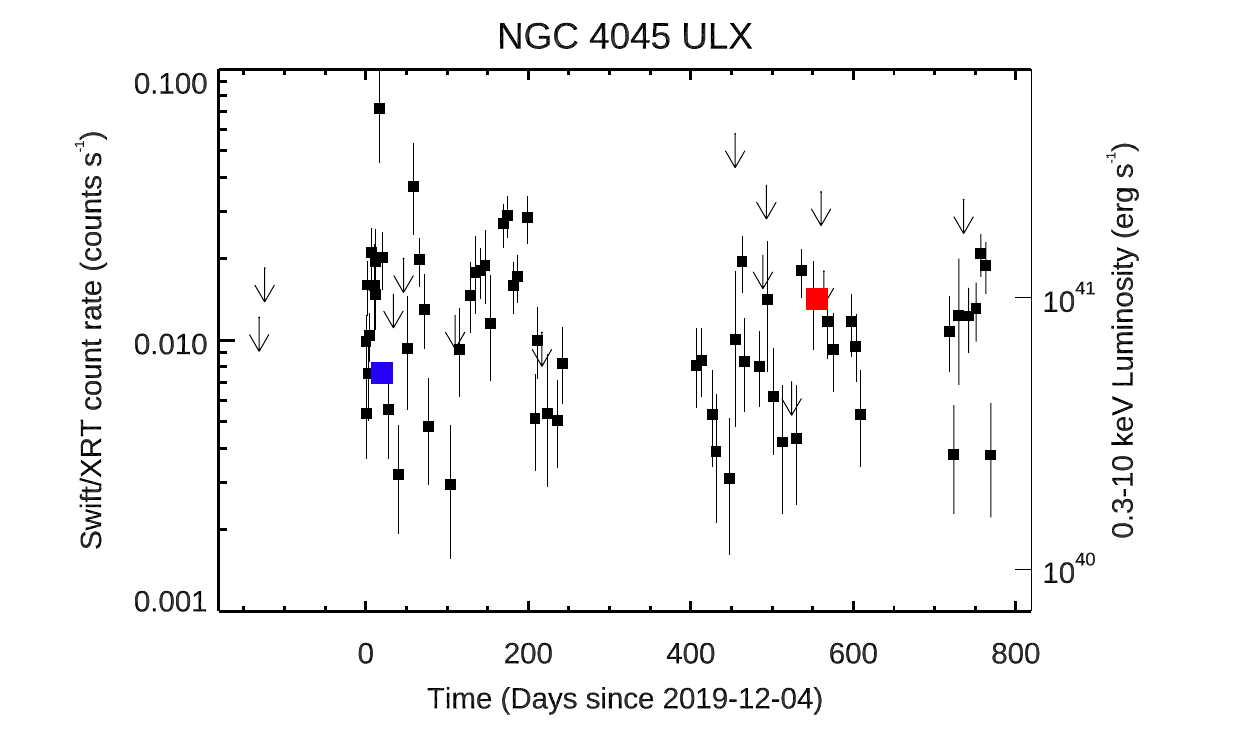}
\end{center}
\caption{\swiftxrt\ lightcurve of the ULX in NGC 4045 (black data points). Upper limits (3$\sigma$) are shown with black arrows. The gaps in the lightcurve are due to \swift\ Sun constraint. The observed 0.3--10 keV luminosity is shown on the right axis and the luminosity of the source when observed with \chandra\ and \nustar\ is shown as a blue square, and when observed with \xmm\ and \nustar\ is shown as a red square.}
\label{fig_ngc4045_ltcrv}
\end{figure}

We continued to monitor the source with already scheduled observations of AT2019wbg and subsequent DDT requests. We used the online tool to extract the \swiftxrt\ lightcurve and spectrum of NGC 4045 ULX from these observations. All products from this tool are fully calibrated and corrected for effects such as pile-up and the bad columns on the CCD. The observations consisted of target IDs 12842, 33099, 89015, 03104800, 03105785, 03105785, and the lightcurve was binned by observation, requiring a minimum detection of 2.5$\sigma$. The tool also fitted an absorbed power-law model to the stacked spectrum which yielded the parameters \nh$=2.6^{+0.9}_{-0.8}\times10^{21}$ \cmsq\ and $\Gamma=1.56^{+0.17}_{-0.16}$ and a count rate to unabsorbed flux conversion factor of $5.69\times10^{-11}$ erg\,cm$^{-2}$\,ct$^{-1}$. We used this to convert the XRT count rate to flux, and assuming the distance of 32.1 Mpc to NGC 4045, convert this to a luminosity. The count rate and luminosity lightcurve is plotted in Figure \ref{fig_ngc4045_ltcrv} and shows that the source is highly variable and also regularly exceeds a luminosity of $10^{41}$ \ergs, classifying it as an HLX.

The hardness ratios from the online tool, defined as the ratio of the 1.5--10 keV count rate to the 0.3--1.5 keV count rate, show potential evidence for spectral evolution from the source. To investigate further, we produce spectra using the online tool for 15 observations during the period 2020-03-26 and 2020-07-19 where HR$>1$, which yielded \nh$=3.8^{+2.2}_{-1.7}\times10^{21}$ \cmsq\ and $\Gamma=1.57^{+0.32}_{-0.30}$ and a count rate to unabsorbed flux conversion factor of $6.27\times10^{-11}$ erg\,cm$^{-2}$\,ct$^{-1}$. For the 8 observations during the period 2020-12-16 and 2021-03-06 where HR$<1$, \nh$=1.2^{+2.3}_{-1.2}\times10^{21}$ \cmsq\ and $\Gamma=2.0^{+0.8}_{-0.7}$ and a count rate to unabsorbed flux conversion factor of $4.22\times10^{-11}$ erg\,cm$^{-2}$\,ct$^{-1}$. The \nh\ and $\Gamma$ measured for these two epochs are consistent with each other within the 90\% uncertainties, therefore we do not find evidence for spectral variations from the \swiftxrt\ data.

\subsection{NuSTAR}

Upon identification of the new ULX, we obtained a director's discretionary time (DDT) observation of the source with \nustar\ \citep{harrison13}. This took place on 2019 December 16 (obsID 90501355002). The source was well detected with a count rate of $5.8\pm0.4\times10^{-3}$ \cntrt\ in FPMA and $5.1\pm0.4\times10^{-3}$ \cntrt\ in FPMB in the 54 ks exposure over the 3--20 keV energy range. Subsequently, we obtained follow-up {\it target of opportunity} (ToO) observations of the source in \nustar\ Cycle 7. The aim was to get a longer exposure with the source at a brighter flux. This took place on 2021 June 11 (obsID 80701507002) and was triggered on the detection of a \swiftxrt\ count rate of $>$0.01 \cntrt\ (Figure \ref{fig_ngc4045_ltcrv}). The source was again well detected, this time with a higher count rate of $9.4\pm0.3\times10^{-3}$ \cntrt\ in FPMA and $9.2\pm0.4\times10^{-3}$ \cntrt\ in FPMB in the 100 ks exposure over the 3--20 keV energy range.

We used {\sc heasoft} v6.28, {\sc nustardas} v2.0.0 and {\sc caldb} v20211115 to analyze the data. We produced cleaned and calibrated events files using {\sc nupipeline} with the default settings on mode 1 data only. We used {\sc nuproducts} to produce spectral data, including source and background spectra, and response files. A circular region with a radius of 40\arcsec\ was used to extract the source spectra. Circular regions with a radius of 90\arcsec\ were used to extract the background spectra, taking care to extract the background from the same chip as the source. For timing analyses, we used the {\sc heasoft} tool {\tt barycorr} to apply a barycentric correction to the event times of arrival, using the default JPL planetary ephemeris DE-200.

\subsection{Chandra}
\label{sec_chandra}

We also obtained a \chandra\ \citep{weisskopf99} DDT observation of the source, which took place on 2019 December 31 (obsID 23106), with ACIS-S at the aimpoint. The source was well detected with a count rate of $2.8\pm0.2\times10^{-2}$ \cntrt\ in the 9.8 ks exposure. We first ran the script {\tt chandra\_repro}, and then extracted the \chandra\ spectra with {\sc specextract} with circular regions, radius 2\arcsec\ for the source and 13\arcsec\ for the background. 

We also used the \chandra\ data to acquire a more precise position for the source. We compiled an X-ray source list of the \chandra\ observation by running {\sc wavdetect} with default parameters on the reprocessed events file, filtered to energies of 0.5--8 keV. This resulted in a source list of 88 X-ray sources. We then cross-matched this with a {\it Gaia} DR3 source list of the region \citep{gaia21}, selecting sources within 1.4\arcsec\ of each other. This identified five joint \chandra/\gaia\ sources. We define the astrometric shifts as the mean difference in RA and Dec between these matched sources, finding $\delta$RA$=+0.03$\arcsec\ and $\delta$Dec$=+0.79$\arcsec. The \chandra\ position of the source, having applied the aforementioned corrections, is R.A. = 12h 02m 42.358s (180.67649\degree), Decl.=+1\degree\ 58\arcmin\ 07.34\arcsec\ (1.9687065\degree, J2000). We adopt the residual offset of 0.50\arcsec\ as our uncertainty, which agrees very well with the \swift\ position (Figure \ref{fig_ngc4045_img}). The only source at other wavelengths catalogued near this position is ULAS J120242.27+015807.6, a UKIDSS-DR9 \citep{lawrence07} $K=17.96$ mag near-infrared source, 1.3\arcsec\ from the \chandra\ position, which is outside our positional error circle. The lack of a multiwavelength counterpart argues against a background AGN as the source of X-rays.

We detect two other X-ray sources that are likely associated with NGC 4045. An X-ray source at R.A. = 12h 02m 42.276s, Decl.=+1\degree\ 58\arcmin\ 36.50\arcsec\ is coincident with the \gaia\ position of the nucleus of the galaxy, 29\arcsec\ from the ULX. It has a count rate of $1.9\pm0.4\times10^{-3}$ \cntrt, corresponding to a flux of $3\times10^{-14}$ \ergcms\ in the 0.5--8 keV band, implying a luminosity of $4\times10^{39}$ \ergs\ at 32 Mpc. This source is likely to be the AGN of NGC 4045, which is at low luminosity.

The other source was at R.A. = 12h 02m 42.799s, Decl.=+1\degree\ 58\arcmin\ 56.98\arcsec\ in the northern spiral arm of the galaxy, 50\arcsec\ from the ULX, and close to, but not coincident with AT2019wbg (Figure \ref{fig_ngc4045_img}). It has a count rate of $2.0\pm1.4\times10^{-4}$ \cntrt, corresponding to a flux of $1.7\times10^{-14}$ \ergcms\ in the 0.3--10 keV band, implying a luminosity of $2\times10^{39}$ \ergs\ at 32 Mpc, and therefore also a ULX.

\subsection{\xmm}

In our \nustar\ Cycle 7 program, we were awarded joint observations with \xmm\ \citep{jansen01}. \xmm\ observed NGC 4045 on 2021 June 12 (obsID 0890610101). We used {\sc xmmsas} v18.0.0 to analyze the data \citep{gabriel04}. We first checked for high background by creating a lightcurve of the events from the entire detector in the 10--12 keV band, finding that the background was low across the entire observation, with less than 0.7 \cntrt\ in this band for the pn detector, and less than 0.35 \cntrt\ in the MOS detectors. Events were selected with {\tt PATTERN$\leq4$} for the pn and {\tt PATTERN$\leq12$} for the MOS. A circular region with a radius of 20\arcsec\ was used to extract the source spectrum. A circular region with a radius of 45\arcsec\ was used to extract the background spectra. Care was taken to extract the background from the same chip as the source, and from the region of low  internal detector Cu K$\alpha$ fluorescence background, located at the center of the detector near to where the source was placed \citep{freyberg04}. Data from the pn and both MOS instruments were extracted in this way. For timing analyses, we used the {\sc xmmsas} tool {\tt barycen} to apply a barycentric correction to the event times of arrival, using the default DE-200 solar ephemeris

The source was well detected with a count rate of $1.57\pm0.01\times10^{-1}$ \cntrt\ in pn, $5.06\pm0.09\times10^{-2}$ \cntrt\ in MOS1, and $5.2\pm0.10\times10^{-2}$ \cntrt\ in MOS2 in the 61 ks exposure over the 0.2--10 keV energy range.

\section{X-ray Spectral Analysis}

We summarize the details of all \nustar, \chandra, and \xmm\ observations of the new HLX in table \ref{tab_obs}. All spectra were grouped with a minimum of one count per bin using the {\sc heasoft} tool {\tt grppha} and fitted in {\sc xspec} \citep{arnaud96}. The C statistic was used for fitting to source spectra with the background subtracted \citep{cash79}. Since the C statistic cannot formally be used when the background is subtracted, {\sc xspec} uses a modified version of the C-statisitc known as the W statistic to account for this. The data are shown in Figure \ref{fig_ngc4045_spec}, top panel. 

\begin{table*}
\centering
\caption{Details of the observations used in this work}
\label{tab_obs}
\begin{center}
\begin{tabular}{c c c c c c c}
\hline
Observatory	& ObsID 		& Start time 		& Instrument	& Exposure$^a$	& Net count rate 			& Flux$^b$  \\
 			& 			&	  (UT)		&			& (ks)			& (\cntrt)					& (\ergcms)  \\
\multirow{2}{*}{\nustar}& \multirow{2}{*}{90501355002}& \multirow{2}{*}{2019-12-16 02:01:09}	& FPMA	& 53.9	& 5.8$\pm0.4\times10^{-3}$	& 4.4$^{+0.3}_{-1.0}\times10^{-13}$  \\
			& 			& 											      	& FPMB	& 53.4	& 5.1$\pm0.4\times10^{-3}$	& 3.9$^{+0.3}_{-1.1}\times10^{-13}$  \\
\chandra\ 		& 23106		& 2019-12-31 05:56:24								& ACIS-S	& 9.8		& 2.8$\pm0.2\times10^{-2}$	& 3.9$^{+0.3}_{-1.0}\times10^{-13}$  \\
\multirow{2}{*}{\nustar}& \multirow{2}{*}{80701507002}& \multirow{2}{*}{2021-06-11 18:51:09}	& FPMA	& 101	& 9.4$\pm0.3\times10^{-3}$	& 7.2$^{+0.3}_{-1.0}\times10^{-13}$  \\
			& 			& 											      	& FPMB	& 99.6	& 9.2$\pm0.4\times10^{-3}$	& 7.3$^{+0.3}_{-1.4}\times10^{-13}$  \\
\multirow{3}{*}{\xmm}& \multirow{3}{*}{0890610101}& \multirow{3}{*}{2021-06-12 04:11:07}	& pn		& 53.1	& 1.6$\pm0.2\times10^{-1}$	& 8.1$^{+0.0}_{-2.0}\times10^{-13}$  \\
			& 			& 											      	& MOS1	& 50.6	& 5.1$\pm0.9\times10^{-2}$	& 8.1$^{+0.2}_{-1.9}\times10^{-13}$  \\
			& 			& 											      	& MOS2	& 61.4	& 5.2$\pm1.0\times10^{-2}$	& 7.9$^{+0.0}_{-1.9}\times10^{-13}$  \\
\hline
\end{tabular}
\tablecomments{$^a$ After filtering.  $^b$ 0.5--8 keV for \chandra, 0.2--10 keV for \xmm, and 3--20 keV for \nustar, observed (absorbed).}
\end{center}
\end{table*}

\subsection{Continuum fitting}

\begin{figure}[h]
\begin{center}
\includegraphics[trim=10 20 20 0, width=90mm]{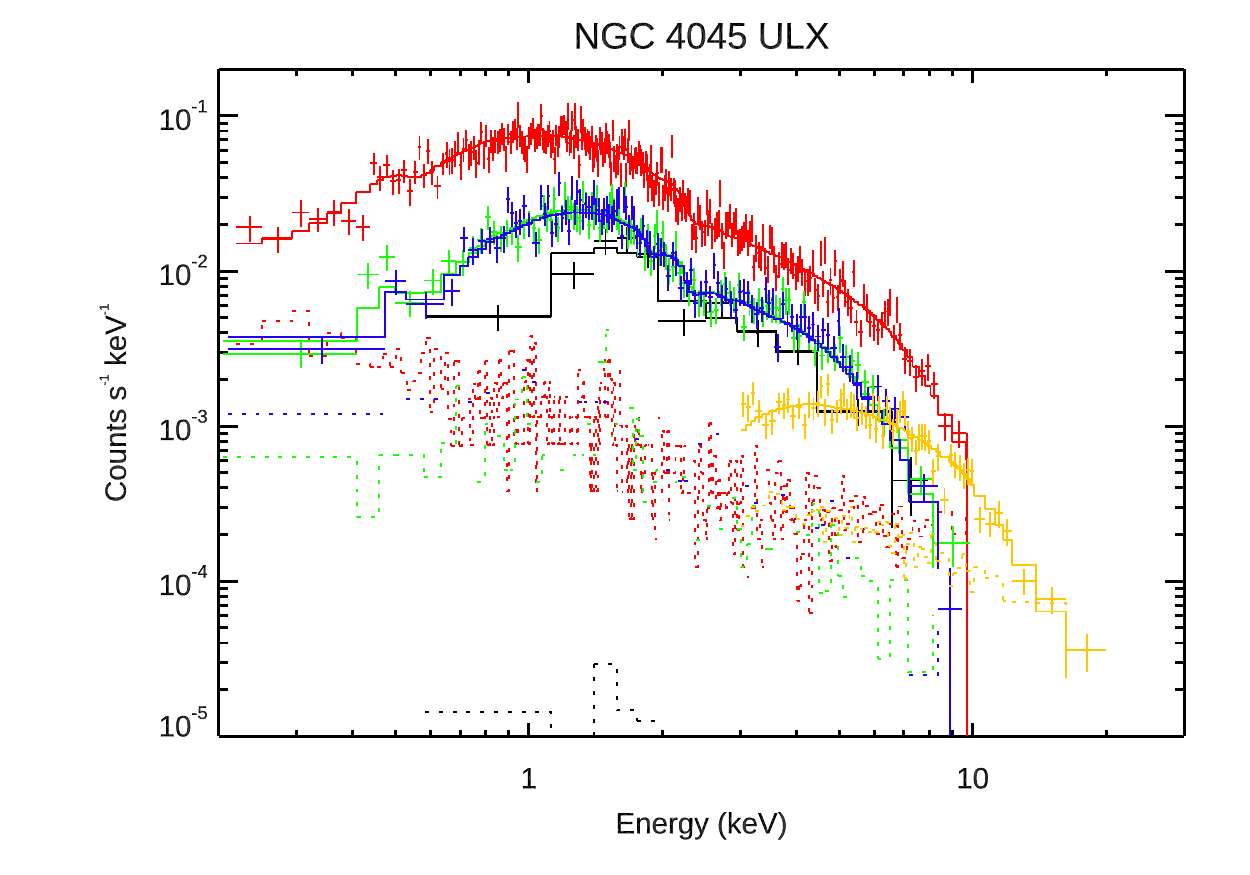}
\includegraphics[trim=10 20 20 0, width=90mm]{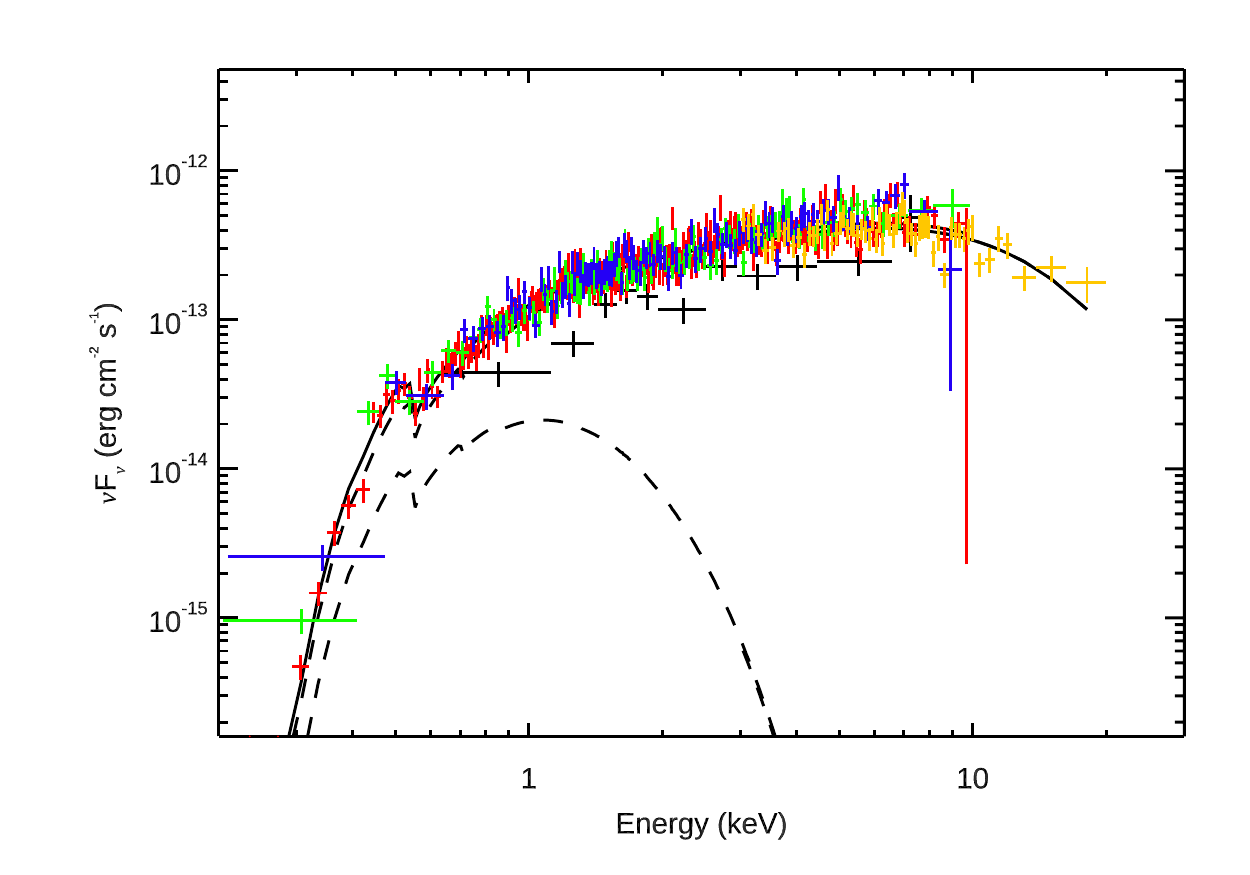}
\includegraphics[trim=10 20 20 0, width=90mm]{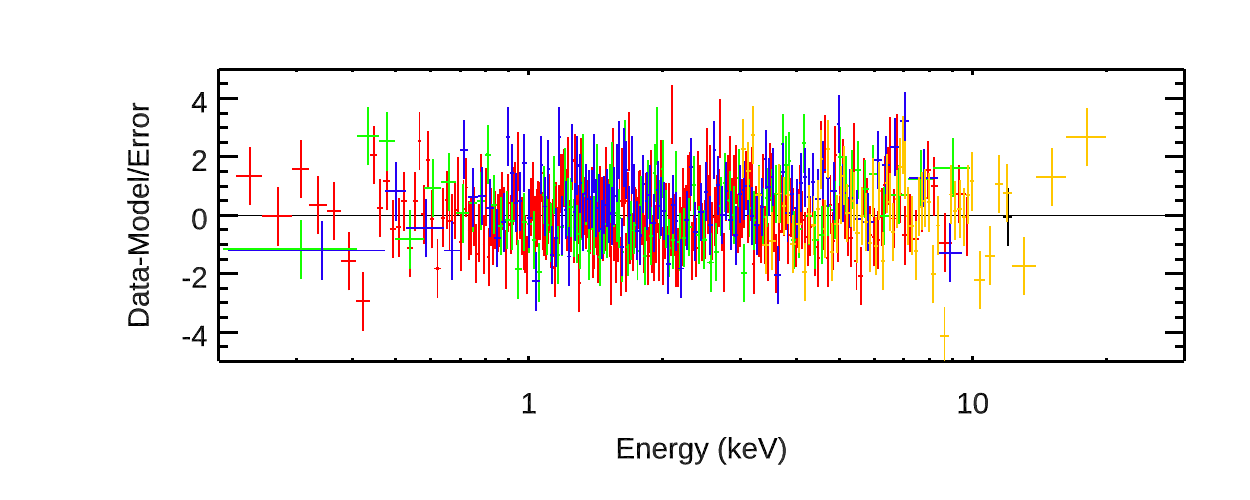}
\end{center}
\caption{\chandra\ (black), \xmm/pn (red), \xmm/MOS1 (green), \xmm/MOS2 (blue), \nustar\ (both epochs FPMA+B, stacked, yellow), spectra of the ULX in NGC 4045. The top panel shows the data with best-fit {\tt tbabs*(diskbb+diskpbb)} model (solid line) and background for each instrument (dotted lines), the middle panel shows the data unfolded through the instrumental responses when assuming the best-fit model, and the bottom panel shows the spectral residuals, rebinned for plotting purposes.}
\label{fig_ngc4045_spec}
\end{figure}

We fit the \chandra, \xmm\ and \nustar\ observations of the source jointly in {\sc xspec} with the use of a {\tt constant} component to account for the flux variability of the source and cross-calibration offsets. We start by fitting a simple absorbed power-law model, {\tt tbabs*powerlaw} with abundances of \cite{anders89}, which yields $C$=4287.24 with 4694 degrees of freedom (DoFs). However the data to model residuals reveal a spectral turn-over, with an energy of $6.7^{+1.1}_{-0.9}$ keV when fitting with the {\tt cutoffpl} model, indicating that this simple model does not represent the data well. Replacing the power-law model with a multi-color disk black body model, with a variable temperature profile index ({\tt diskpbb}), improves the fit substantially to $C$=4121.29 with 4693 DoFs. While this model appears to represent the data well, ULX spectra often exhibit two disk-like components, a cooler one which may come from the outer regions of an accretion disk or the photosphere of an outflow \citep{qiu21}, and a hotter component, which may originate from the inner regions of the accretion disk \citep{walton18c}, an accretion curtain \citep{mushtukov17}, or Compton up-scattering \citep{gladstone09}. We therefore tried the addition of a second cooler disk-like component, {\tt diskbb}, which gives $C$=4117.17 with 4691 DoFs, only a minor improvement to the fit. This is probably due to the relatively high absorption in the system and the dominance of the hotter component. However, we keep it for comparison to other ULXs. The fit with this model is shown in Figure \ref{fig_ngc4045_spec}, middle panel, and we list the spectral parameters and their uncertainties in Table \ref{tab_specpar}.

\begin{table}
\centering
\caption{Joint X-ray continuum spectral fitting results.}
\label{tab_specpar}
\begin{center}
\begin{tabular}{l l}
\hline
\multicolumn{2}{c}{\tt tbabs}\\
\nh & 2.1$^{+0.9}_{-0.4}\times10^{21}$ \cmsq\ \\
\multicolumn{2}{c}{\tt diskbb}\\
$T_{\rm in}$ & 0.27$^{+0.19}_{-0.08}$ keV \\
Normalization & 0.74$^{+11}_{-0.71}$ \\
\multicolumn{2}{c}{\tt diskpbb}\\
$T_{\rm in}$ & 3.27$^{+0.38}_{-0.32}$ keV \\
$p$ & 0.60$^{+0.04}_{-0.03}$ \\
Normalization & 2.0$^{+1.7}_{-0.9}\times10^{-4}$ \\
\multicolumn{2}{c}{\tt constant}\\
$C_{\rm FPMA, 2019}$ & 0.67$\pm0.07$  \\
$C_{\rm FPMB, 2019}$ & 0.60$^{+0.07}_{-0.08}$  \\
$C_{\rm ACIS}$ & 0.56$^{+0.06}_{-0.05}$ \\
$C_{\rm FPMA, 2021}$ & 1.08$\pm0.07$  \\
$C_{\rm FPMB, 2021}$ & 1.09$^{+0.07}_{-0.08}$  \\
$C_{\rm pn}$ & 1.00 (fixed)\\
$C_{\rm MOS1}$ & 1.01$\pm0.04$\\
$C_{\rm MOS2}$ & 0.98$\pm0.04$\\
\fx\ (0.2--20 keV)$^a$ & $9.8^{+0.1}_{-1.9}\times10^{-13}$ \ergcms\ \\
\lx\ (0.001--100 keV)$^b$ & $1.5^{+0.6}_{-0.1}\times10^{41}$ \ergs\ \\
C-statistic & 4117.17 \\
DoFs & 4691 \\ 
\hline
\end{tabular}
\tablecomments{Results from the continuum spectral fitting of the \nustar, \chandra\ and \xmm\ data. $^a${\it observed}, $^b${\it unabsorbed}.}
\end{center}
\end{table}

We tested for spectral variability between the 2019 \chandra+\nustar\ observations, and the 2021 \xmm+\nustar\ observations by allowing the parameters to vary in the fit one by one, and calculating their 90\% uncertainties. We found no evidence for spectral variability, finding that the parameters were consistent with each other within the 90\% confidence uncertainties between epochs.

\subsection{Absorption line fitting}

\begin{figure}[h]
\begin{center}
\includegraphics[trim=20 20 20 0, width=90mm]{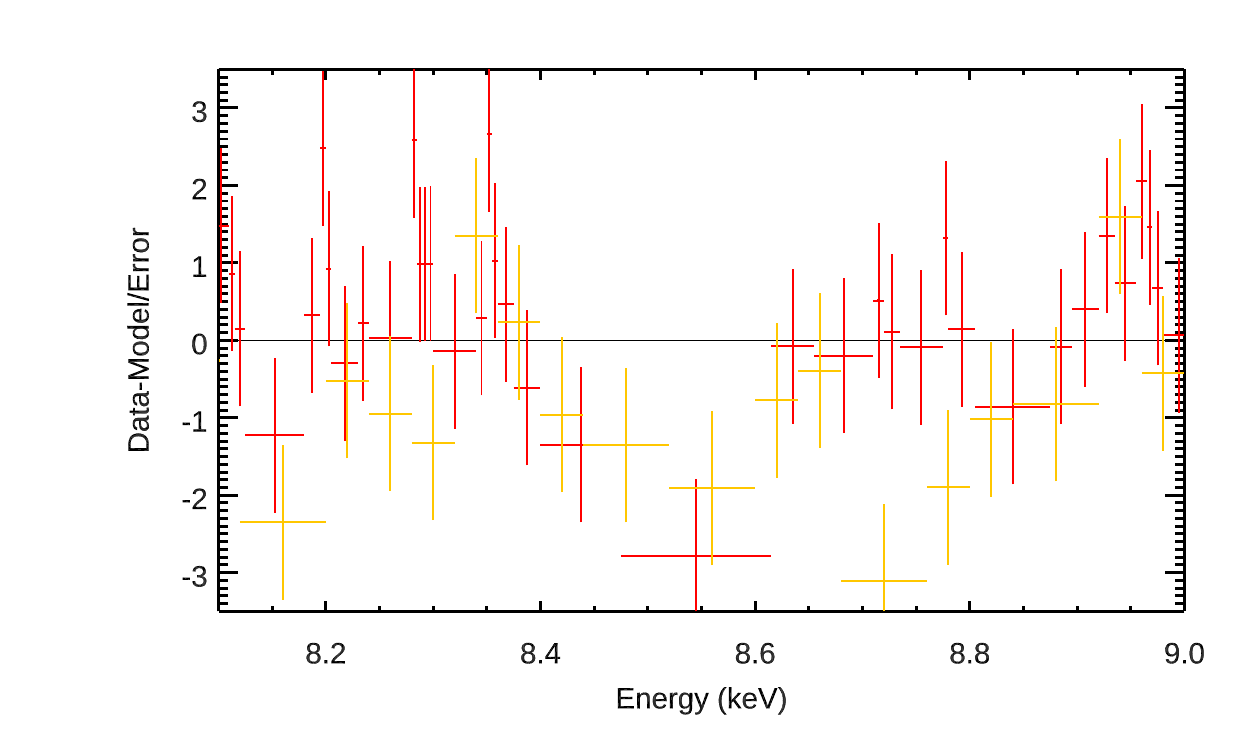}
\end{center}
\caption{Spectral residuals of the best-fit continuum model in the 8--9 keV band, where a deficit of counts can be seen at $\sim$8.5 keV. \xmm/pn (red), \nustar\ (both epochs FPMA+B, stacked, yellow) are plotted, rebinned for plotting purposes. \chandra\ and MOS data are omitted for clarity.}\label{fig_ngc4045_delc}
\end{figure}

While the spectral residuals do not indicate any other continuum model components, we noted a deficit of counts in the 8--9 keV band that could be an absorption line (Figure \ref{fig_ngc4045_delc}). To test this hypothesis, we add a Gaussian absorption component {\tt gauss} to the fit. This yields $C$=4099.62 with 4688 degrees of freedom (DoFs), an improvement to the fit of $\Delta C=-17.5$ with the loss of 3 DoFs. The contributions to the observed change in $C$ are $\Delta C=-3.7$ from FPMA+B (2019); $\Delta C=-6.9$ from FPMA+B (2021); $\Delta C=-6.2$ from pn; and -0.5 from MOSs. This shows that the improvement in the fit is not driven by a single instrument, or observing epoch, indicating that the absorption feature is neither instrumental nor transient. 

We first determine the false alarm rate (FAR), the probability that a change in $C$ with this magnitude is a statistical fluctuation. We do this with simulations using the {\tt fakeit} command in {\sc xspec}. We generate 6000 simulated spectra from each instrument based on the continuum {\tt tbabs*(diskbb+diskpbb)} model only and the observed background and instrumental responses. We generate unbinned data from these and group them with a minimum of 1 count per bin as done with the real data. 

\begin{figure}[h]
\begin{center}
\includegraphics[trim=20 20 20 0, width=90mm]{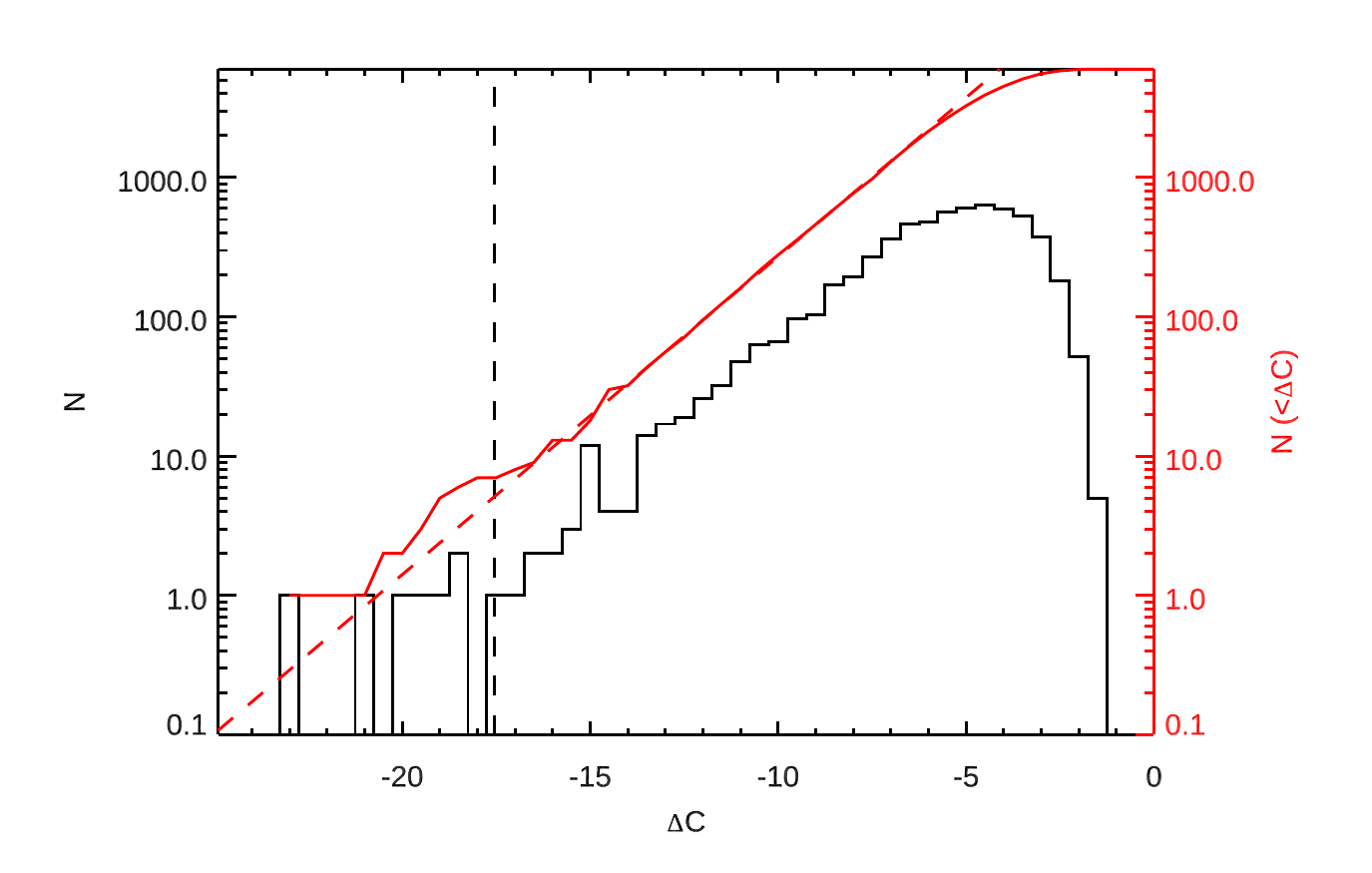}
\end{center}
\caption{Results of the $5000$ spectral simulations used to determine the significance of the absorption line at 8.58 keV. The black histogram shows the distribution of $\Delta C$ produced by the simulations, and the dashed black line shows the observed value. Only 6 simulated values exceed the $\Delta C$ observed. The solid red line shows the cumulative distribution of the $\Delta C$ produced by the simulations, and the red dashed line is a fit to this distribution, which implies the number of simulated spectra that produce a $\Delta C$ as large as that observed is 5.1.}
\label{fig_dcstat}
\end{figure}

We then refit the simulated spectra with the continuum model and add the {\tt gauss} component. Since we cannot visually search each simulated spectrum for residuals as done for the real data, in order to ensure we find the strongest residual in each simulated spectrum, we perform a Gaussian line scan using the {\tt steppar} command in {\sc xspec} to search over the line energy in the 2--10 keV energy range in 80 equally and linearly spaced steps. This procedure fits the spectrum at each step, with the line energy and width fixed, but the normalization free to vary. The line width is fixed at 0.1 keV to simulate the unresolved nature of the real feature. For each simulated spectrum set, we note the largest change in $C$ when carrying out the line search. We plot the distribution of these maximum $C$ values in Figure \ref{fig_dcstat}. Only 7 simulated spectrum sets produce a $\Delta C$ as large as that observed, implying a FAR of 1.2$\times10^{-3}$.

Since the number of simulated $\Delta C$ is very small at large negative values, this FAR is subject to small-number statistics. We also plot the cumulative $\Delta C$ distribution Figure \ref{fig_dcstat}, which is basically the FAR for a given $\Delta C$ when normalized by the number of simulations. The $\Delta C$ and cumulative $\Delta C$ distributions clearly take power-law form for $\Delta C > -15$, below which small number statistics skew the distributions. We then fit the $\Delta C$ distribution analytically for $\Delta C > -15$ and extrapolate to more negative values to determine the precise FAR. Fitting the cumulative $\Delta C$ distribution with a power-law yields a power-law index of $0.090$. This predicts the number of simulations where $\Delta C < -17.5$ is 5.1, which when divided by 6000 implies the FAR is $8.5\times10^{-4}$. This is equivalent to $>3.3\sigma$, which confirms that the absorption line is significantly detected. The implied 3-$\sigma$ $\Delta C$ threshold is -15.2.

\begin{figure}[h]
\begin{center}
\includegraphics[trim=20 20 20 0, width=90mm]{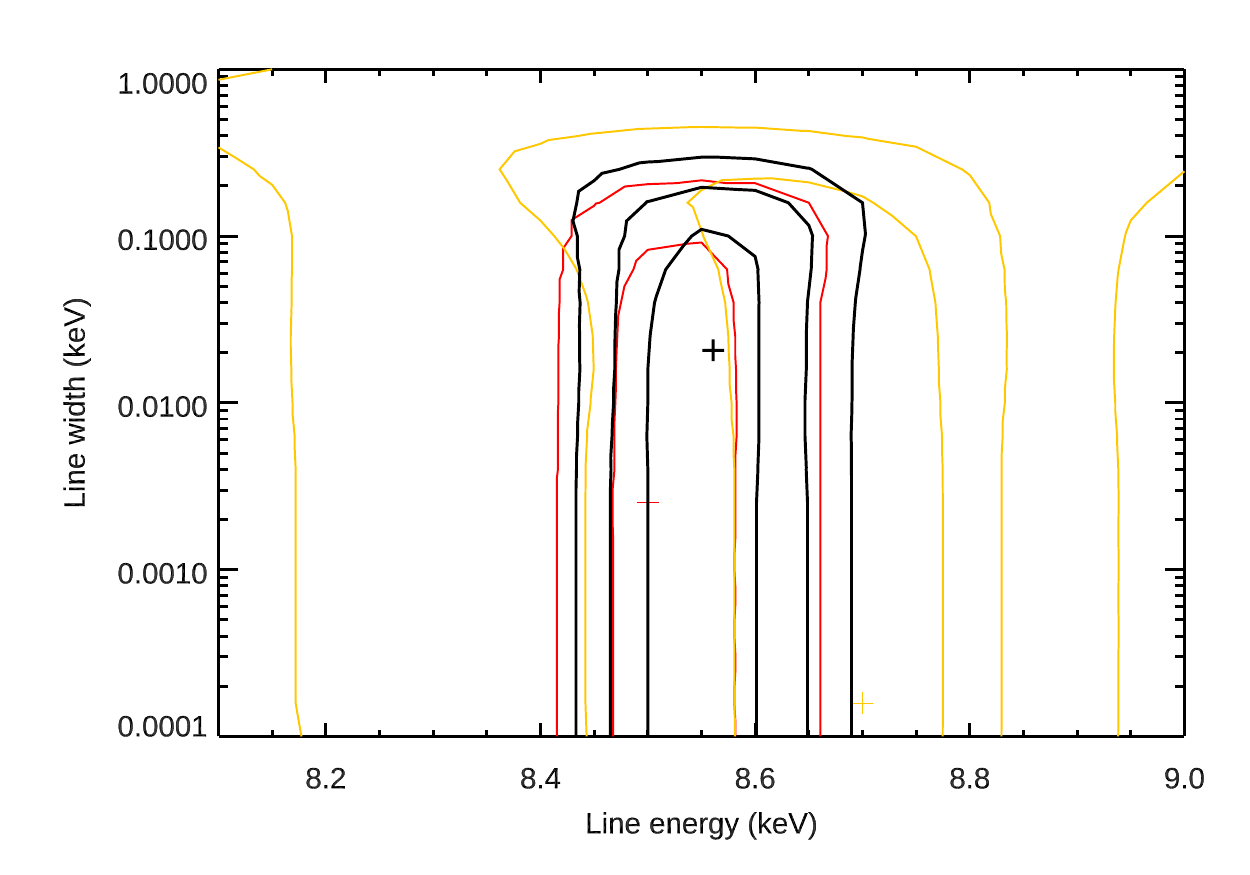}
\end{center}
\caption{1, 2 and 3$\sigma$ $C$-statistic contours of the energy and width of the Gaussian absorption line for all instruments together (black), for \xmm/pn (red) and for \nustar/FPMA+B (yellow, 2019 and 2021 epochs) when fitted individually.}
\label{fig_linee_sig_con}
\end{figure}

The best-fit parameters of the Gaussian absorption line for all instruments together are a line energy $E=8.56^{+0.05}_{-0.11}$ keV, width $\sigma<0.2$ keV and normalization $K=-7.4^{+2.6}_{-2.4}\times10^{-7}$ photons cm$^{-2}$ s$^{-1}$ with an equivalent width EW$=-0.22^{+0.08}_{-0.09}$ keV. We also tested for spectral variability of the absorption line between observational epochs, but found no evidence for it. We show the contour plot of the energy and width of the line in Figure \ref{fig_linee_sig_con} showing that the width of the line is unconstrained at the lower end due to being close to the instrumental energy resolution of \xmm/pn. This also shows the constraints provided by the pn, and two FPMA epochs showing they are all consistent with each other within their 1-$\sigma$ confidence levels.

We also test the possibility that the 8.56 keV feature is an absorption edge rather than a line doing so using the {\tt edge} model in {\sc xspec}. Adding this parameter to the continuum model yields $C$=4107.98 with 4689 degrees of freedom (DoFs), an improvement to the fit of $\Delta C=-9.2$ with the loss of 2 DoFs. This is not as great an improvement as the line case, despite one fewer DoFs, therefore we conclude that the feature is more line like than edge like.

\subsection{Potential identification as absorption by a highly ionized outflow}
\label{sec_pion}

The 8.56 keV absorption line is potentially associated with a highly ionized outflow produced by a high-velocity disk wind. In an attempt to understand the properties of the potential ionized outflow, we fit the \nustar\ and \xmm\ spectra with the PION spectral model within the \textsc{spex} package \citep{kaastra96}. PION \citep[e.g.][]{miller15} determines the ionizing balance and the absorption line strengths from the spectral energy distribution (SED) of the currently loaded \textsc{spex} continuum model. If the data quality is sufficient, the model might allow us to measure the ionization parameter ($\xi$), column density (\nh) as well as the projected velocity ($v$) of the outflow. Then it would be possible to estimate the outflow energetics.

Since PION is based in \textsc{spex}, the continuum models used here are slightly different to those used above in \textsc{xspec}. The hotter spectral component is modeled with an MBB component (a blackbody modified by coherent Compton scattering), with a temperature of $\sim3$ keV. The cooler spectral component is modeled by a simple blackbody with a temperature of 0.3-0.4 keV. Both are obscured by the HOT component, which describes the transmission through a layer of a collisionally ionized plasma, with a temperature of 0.5 eV, simulating nearly neutral ISM plasma in collisional equilibrium, and assumes the abundances of \cite{lodders09}. The column density is \nh$\sim8 \times 10^{20}$ cm$^{-2}$, and describes the interstellar absorption in both our Galaxy and NGC 4045. We do not find evidence for any additional continuum components.

This continuum model is used to fit all the \xmm\ and \nustar\ spectra, converted to \textsc{spex} file format using the Trafo procedure, in the appropriate energy ranges, binned to at least one count per spectral bin. The component parameters including normalizations are tied across the observations, with only the overall model normalization being left free to vary. The resulting continuum model results in $C=$4157.69 with 4736 DoF.

Following the continuum fit, we applied the PION model. The fit statistics improved to $C=4141.83$ with the loss of 4 DoFs, for a fit improvement of $\Delta C=-15.86$. We find a column density \nh=$2^{+0.0}_{-0.7} \times 10^{24}$ \cmsq, an ionization parameter log($\xi$/erg cm s$^{-1}$)=$4.70\pm0.14$, a velocity width of $2900^{+2200}_{-1200}$ \kms\ and a projected systematic velocity of $61200^{+1700}_{-1500}$ \kms\ (relativistically corrected). Unfortunately, the column density of the PION model is pegged to our upper limit of $2\times10^{24}$ \cmsq. This is a very high value, exceeding the UFO column densities measured in both ULXs and AGNs \citep[e.g.][]{tombesi11,pinto16,pinto21}. An outflow of such high column density would plausibly produce a strong associated emission signature such as a P-Cygni feature with a strong emission component, but this is not observed in the current spectra. However, the presence of a P-Cygni profile depends on the exact geometry and solid angle of the wind responsible for the Fe K feature. Additionally, such a high column density requires a very large unabsorbed 0.3--30 keV X-ray luminosity of $7 \times 10^{41}$ \ergs, higher than that observed and on the very upper end of the ULX luminosity function.

Therefore we do not consider such a high \nh\ value to be trustworthy. The poor data quality could be the reason for the unrealistic value. At such high ionization levels, the ionization parameter and thus the column density are determined from the ratio of the \fexxv\ (rest-frame energy of 6.67 keV) and \fexxvi\ (6.97 keV) absorption line strengths. The poor data quality could result in very inaccurate measurement of this ratio. Additionally, by increasing the column density (and thus the ionization parameter), the PION model increases the absorption strengths of the Fe K lines while decreasing the strengths of the lower energy lines such as \oviii, \nex, \sixiv\ and \sxvi. These lower energies ($<$3 keV) are poorly resolved with the EPIC detectors (and outside the \nustar\ range) but it appears that no (even weak) absorption residuals are present at the appropriate energies of the other elements. Thus PION chooses to reduce the line strengths as much as possible. Naturally, no additional (lower energy) absorption features would be expected in case that the residual is due to noise or is a cyclotron resonance feature, but this could still be a spectral resolution issue (at the present data quality) in the soft X-ray band. The column density could also be reduced by increasing the Fe abundance, however the abundance value is highly degenerate with the absorber column density and we could not obtain a reasonable fit.

Further, longer \xmm\ and \nustar\ observations of the source might improve the data quality, leading to a more confident UFO parameter measurement allowing us to estimate the wind energetics. An RGS grating detection of the blueshifted \oviii\ feature (expected around 0.8 keV) would lead to a much more trustworthy plasma parameter measurement.

The column density is strongly degenerate with the ionization parameter. We can therefore fix $\log \xi$ to a more realistic value and recover the column density. Assuming log($\xi$/erg cm s$^{-1}$)=3.92, the best-fitting ionization parameter of the UFO in the neutron star ULX NGC 300 ULX-1 \citep{kosec18}, we recover a column density of  $2.2^{+0.9}_{-0.7}\times 10^{23}$ \cmsq. This is a much more realistic \nh\ value, comparable with the one measured in NGC 300 ULX-1 (best-fitting column density of $1.2^{+1.9}_{-0.6} \times10^{23}$ \cmsq). However, the addition of a PION component of such parameters results in poorer fit improvement of just $\Delta C=-13.7$ over the baseline continuum model.

\subsection{Potential identification as a cyclotron resonance scattering feature}

Above we explored the identification of the 8.56 keV absorption line with absorption by highly ionized iron. Here we explore the possibility that the absorption line is a cyclotron resonance scattering feature (CRSF), produced by the interaction of X-ray photons with a strong magnetic field. This would naturally identify the compact object as a neutron star, since black holes are not capable of producing such strong magnetic fields.

To test this possibility, we trial the use of the {\sc xspec} model {\tt cyclabs} \citep{mihara90,makishima90}. This model utilizes two Lorenzian functions to represent the fundamental and first harmonic lines, with the energy of the harmonic fixed at twice the energy of the fundamental. We assume the 8.56 keV line to be the fundamental and set the width of the harmonic line to be the same as the fundamental. Adding and fitting this model to our continuum model yields $C=4101.73$ with 4687 DoFs, slightly worse that the single Gaussian absorption line. 

The energy of the fundamental is $E=8.56^{+0.09}_{-0.08}$ keV with a width of $W<0.12$ keV and an optical depth of $1\leq D_{\rm 0} \leq 6\times10^3$. The energy of the harmonic is implied to 17.1 keV, however the depth is unconstrained with an upper limit of $D_{\rm 2} \leq 6\times10^4$. This loose constraint is because the number of counts and signal to noise of the \nustar\ data at this energy is low. Determining the presence of a harmonic line is key to identifying the line as a CRSF, especially when the fundamental line has an energy consistent with atomic absorption. However, we are unable to do this with the current data.

Alternatively, if we assume the 8.6 keV line is the harmonic, and the fundamental is at 4.3 keV, then we can place an upper limit of 180 on the depth of the fundamental, compared with a lower limit of 660 on the harmonic, i.e. the depth of the harmonic is constrained to be $>3$ times that of the fundamental.

\subsection{Potential AGN contamination}

As described in Section \ref{sec_chandra}, we identified a low-luminosity AGN in NGC 4045 in the \chandra\ data. The AGN was 29\arcsec\ from the ULX, and therefore potentially contaminates the \xmm\ and \nustar\ PSFs of the source. We extracted the \chandra\ spectrum of the AGN and fitted it with an absorbed power-law, which yielded \nh$<3\times10^{21}$ \cmsq\ and $\Gamma=2.9^{+1.2}_{-0.9}$ with a 0.5--8 keV flux of 3$\times10^{-14}$ \ergcms, less than 10\% of the flux of the ULX. The spectrum is also softer than the ULX, contributing less than 1\% of the flux in the 8--9 keV band where the absorption line was identified. We therefore rule out AGN contamination as a source of error in the absorption line analysis. 

\section{Pulsation search}

We searched for pulsations in the \xmm\ and \nustar\ data using the \texttt{HENDRICS} command line tools, which are based on the \texttt{Stingray} package \citep{bachetti18, huppenkothen19}. In particular, we used the \texttt{HENaccelsearch} to search for pulsations between 0.006--2\,Hz (0.5 to 167\,s), a range in which most ULX pulsars are found. We could not find any significant pulsations, neither in \xmm\ nor in the \nustar\ data in this range, likely due to too few counts collected.

We then proceeded to calculate upper limits on the pulsed fraction of any possible pulsation between 0.1--1\,Hz. To do that, we simulated event files through a Poisson process with the same number of events, exposure time, and GTI windows, but injected a sinusoidal pulsation profile with a given pulsed fraction.  For each pulsed fraction we performed 60 simulations and measured how often we could recover the pulsations at the 99\% significance limit and determine the upper limit on the pulsed fraction as where this is the case for 90\% of all simulations. Based on these simulations we find $\text{PF}<15\%$ for \xmm\ and $\text{PF}<35\%$ for \nustar. The pulsed fraction for ULX pulsars is $\sim$10--30\%, increasing with energy \citep[e.g.][]{bachetti14,fuerst16,israel17}.

\section{Discussion}

\subsection{The broadband X-ray spectrum of NGC 4045 ULX}

The current sample of ULXs with high-quality, broadband X-ray spectra is relatively small, only $\sim$10 sources \citep{walton18c}. We can now add NGC 4045 ULX to that sample. The broadband X-ray spectrum of NGC 4045 ULX can be well described by two disk-like components similar to other ULXs \citep{middleton15a,pintore17,koliopanos17,walton18c}. One of these is potentially associated with the hot, large-scale-height inner flow, and the other a cooler component, perhaps associated with the outer part of the disk or the photosphere of an outflow. Furthermore, a high-energy tail is often detected, possibly associated with the accretion column of a pulsar, or Comptonized emission. As presented by \cite{walton18c}, this component can often be seen as well, which can be modeled with a {\tt cutoffpl} model. If we add this model to our fit, with $\Gamma=0.5$ and $E_{\rm cut}$=8.1 keV fixed, the fit improves slightly to $C=4112.93$ with 4690 DoFs. In the sample of \cite{walton18c} fitted with the three component model, the temperature of the cool component for ULXs ranges from 0.2--0.5 keV and the hotter component has a temperature range of 1.2--3 keV. We find for NGC 4045 ULX that cool disk-like component has a temperature of 0.40 keV, while the hot one has a temperature of 2.1 keV for NGC 4045 ULX, completely consistent with other ULXs. \cite{walton18c} also noted that the ratio of the temperatures of these two components for ULX pulsars in their sample was $\sim3$, while the other ULXs had a temperature ratio of $\sim8$. For the NGC 4045 ULX, the ratio is 5, in between the wider ULX population and ULX pulsars. 

\subsection{An ultrafast outflow from NGC 4045 ULX}

If the 8.56 keV absorption line is from highly ionized iron, it is either from \fexxv\ with a rest-frame energy of 6.67 keV or from \fexxvi\ with a rest-frame energy of 6.98 keV, making the outflow velocity in our line of sight $v=$0.23--0.28$c$, similar to the velocities of the outflows seen in other ULXs \citep[0.1--0.3c, e.g.][]{pinto16}. Unfortunately, we could not model the parameters of the outflow well from this single line. The EW is high, $-0.22^{+0.08}_{-0.09}$ keV, several times stronger than the UFO seen in the Fe K band for NGC 1313 X-1 \citep[-61 eV,][]{walton16}, and requires a high ionization state, column density, or both.

\subsection{A neutron star powering NGC 4045 ULX?}

If the 8.56 keV absorption line is produced by cyclotron resonance scattering of X-ray photons by charged particles in the presence of a powerful magnetic field, this would imply that NGC 4045 ULX is powered by a neutron star, since black holes cannot produce such strong magnetic fields.
If the charged particles are electrons, the transition energy, $\Delta E$, is $11.6(1+z)^{-1}(B/10^{12}$ G)~keV, where $z=(1-2GM/r_{cyc}c^2)^{-1/2}-1$ ($r_{cyc}$ is the radius at which the CRSF forms, assumed to be the surface of the neutron star) is the gravitational redshift and is $\sim0.25$ for the emission from the surface of a typical neutron star. The 8.56~keV line that we have detected would therefore imply $B=7(1+z)\times10^{11}$ G. If the charged particles are protons, $\Delta E=6.3(1+z)^{-1}(B/10^{15}$ G)~keV, thus, interpreting our observed line as a proton CRSFs would imply a very high magnetic field strength of $1.4(1+z)\times10^{15}$ G.

For Galactic pulsars with known electron CRSFs, the features are typically broad with Gaussian line widths of order $\sim1$~keV, and are seen mostly at energies above 10~keV, giving broadening ratios, $\sigma/E\sim0.1$ \citep{tsygankov06,jaisawal16}. Protons on the other hand are more massive and should produce narrower lines. The few proton CRSF observed to date were indeed narrow ($\sigma<0.4$~keV) and at energies below 10~keV \citep{ibrahim02}, giving broadening ratios of $\sigma/E<0.1$. The broadening ratio of the line we have observed is $<$0.014 and therefore more comparable to the previously reported proton CRSFs. However, NGC 4045 ULX is the most luminous source with a potential CRSF identified so far, so drawing a connection to lower luminosity neutron star systems may be tenuous.

The detection of a harmonic line would be key in confirming the CRSF scenario, and for identifying the charged particles as electrons or protons. In Galactic pulsars, the fundamental line is often observed to be weaker than the harmonic line due to photon spawning, caused by transitions from high to low Landau levels that produce photons at the energy of the fundamental line \citep{araya99}. This effect is strongest for electrons and for hard X-ray spectra. The harmonic line is expected at 17 keV, and thus only currently observable with \nustar.

Finally, while no pulsations have yet been detected from NGC 4045 ULX, which would also unambiguously identify a neutron star powering it, we cannot rule out pulsations, since the upper limits on the pulsed fraction we derive are above those of the typical ULX pulsar.

\subsection{Summary and Conclusions}

We have identified a new hyperluminous X-ray source (\lx$>10^{41}$ \ergs) in the galaxy NGC 4045 at a distance of 32 Mpc. We have presented \swift, \nustar, \chandra\ and \xmm\ observations of the source which show the broadband spectrum being very similar to other ULXs. We have also found an absorption line significantly detected ($>3.3\sigma$) at 8.56 keV which appears most likely to be the signature of a highly ionized ultra fast outflow. However, the large equivalent width of the line (EW$=-0.22^{+0.08}_{-0.09}$ keV) paired with the lack of other absorption lines detected requires a high column density and ionization parameter in which case a P-Cygni line profile would be expected, combined with a high luminosity. An alternative explanation is that the line is due to a cyclotron resonance scattering feature (CRSF), produced by the interaction of X-ray photons with a powerful magnetic field. Further observations with high spectral resolution at low energies to detect other signatures of the outflow, or deeper observations at $>10$ keV to detect the harmonic line of the CRSF are needed to differentiate between these scenarios.

\facilities{Swift, NuSTAR, CXO, XMM} 

\software{{\sc NuSTARDAS}, {\sc XMMSAS} \citep{gabriel04}, {\sc XSPEC} \citep{arnaud96}, \textsc{spex}, \citep{kaastra96}, {\tt hendrics} \citep{bachetti15}}

\acknowledgements{
We thank the anonymous referee for their review of this manuscript and for their constructive input.

We wish to thank the \swift\ PI, Brad Cenko for approving the target of opportunity requests we made to observe NGC 4045 ULX, as well as the rest of the \swift\ team for carrying them out. We also acknowledge the use of public data from the \swift\ data archive. This work made use of data supplied by the UK Swift Science Data Centre at the University of Leicester. 

We also wish to thank the \nustar\ PI, Fiona Harrison, for approving the DDT request we made to observe NGC 4045 ULX, as well as the \nustar\ SOC for carrying out the observation. This work was also supported under NASA Contract No. NNG08FD60C. \nustar\ is a project led by the California Institute of Technology, managed by the Jet Propulsion Laboratory, and funded by the National Aeronautics and Space Administration. This research has made use of the NuSTAR Data Analysis Software (NuSTARDAS) jointly developed by the ASI Science Data Center (ASDC, Italy) and the California Institute of Technology (USA).

In addition, we wish to thank Belinda Wilkes, Director of the Chandra X-ray Center for approving the DDT request to observe NGC 4045 ULX and the \chandra\ team for carrying out the observation.

Finally we wish to thank Timothy Kallman for help in generating XSTAR models for XSPEC that were used in the analysis.

This work was also based on observations obtained with XMM-Newton, an ESA science mission with instruments and contributions directly funded by ESA Member States and NASA

This research has made use of data and/or software provided by the High Energy Astrophysics Science Archive Research Center (HEASARC), which is a service of the Astrophysics Science Division at NASA/GSFC.

The work of DS was carried out at the Jet Propulsion Laboratory, California Institute of Technology, under a contract with NASA. }

\bibliography{NGC 4045 ULX.bbl}

\end{document}